# Universal coupler for bulk whispering gallery mode resonators based on silicon subwavelength grating waveguides

D. Farnesi[1], S. Pelli[1], S. Soria[1], G. Nunzi Conti[1], X. Le Roux[2], M. Montesinos Ballester[2], L. Vivien[2], P. Cheben[3,4] and C. Alonso-Ramos[2]

[1] CNR-IFAC Institute of Applied Physics "N. Carrara", 50019 Sesto Fiorentino, Italy
[2] Centre de Nanosciences et de Nanotechnologies, CNRS, Université Paris-Sud, Université Paris-Saclay, Palaiseau 91120, France
[3] National Research Council Canada, 1200 Montreal Road, Bldg. M50, Ottawa, Ontario K1A 0R6, Canada
[4] Center for Research in Photonics, University of Ottawa, Ottawa, Ontario K1N 6N5, Canada

Optical microresonators are of paramount importance in photonic circuits requiring fine spectral filtering or resonant light recirculation. Key performance metrics improve with increasing resonance quality factor (Q) across all applications. The performance of silicon photonic circuits is often hampered by the low-quality factor of planar silicon microresonators, typically of $Q \sim 10^{4\text{-}5}$. On the other hand, bulk whispering gallery mode resonators provide a wide range of materials with intriguing optical properties and exceptionally high resonant quality factors $Q>10^7$. However, the efficient coupling between bulk resonators and planar Si photonic waveguides is considered challenging, if not impossible, due to remarkably large mismatch in size and refractive index. Here, we show an efficient method to couple bulk resonators and Si waveguides based on subwavelength metamaterial engineering of silicon. Based on this approach, we experimentally demonstrate coupling between 220-nm-thick Si waveguides and bulk microresonators made of silica, lithium niobate and calcium fluoride with diameters in the 0.3-3.5 mm range, achieving high coupling efficiency of 75-99% and exceptional Q of $10^6$-$10^7$. These results open a new route for the heterogeneous integration of bulk resonators and silicon photonic circuits, with great potential for applications in sensing, microwave-photonics, and quantum photonics, to name a few.

Optical microresonators provide key functionalities like ultra-fine spectral filtering, lasers with high spectral purity, high-efficiency frequency comb generation, and high optomechanical coupling. The performance of the resonators is typically assessed by their quality factor, Q, which is inversely proportional to the fractional power loss per round trip. Key performance metrics improve with increasing Q across all applications. For example, power consumption and phase and intensity noise in resonator-based optical sources, scale inversely with the square of Q. At the same time, high Q improves precision in resolving the resonance wavelength, which is key for sensing [1] or for frequency stabilization [2].

Silicon photonics provides a unique potential for large volume production of optoelectronic circuits. Thus, Si photonics is considered a key technology for the development of emerging applications in sensing and communications. However, the performance of silicon photonic circuits is in often limited by the comparatively low-quality factor of planar Si microresonators, in the $10^4$ - $10^6$ range. In addition, Si has some intrinsic physical limits that hamper the implementation of advanced functionalities. Namely, strong two-photon absorption in the near-infrared, limiting nonlinear applications, and no Pockels effect, compromising the efficiency of optoelectronic circuits. On the other hand, bulk whispering gallery mode (WGM) optical resonators like spheres and disks provide a wide range of materials with remarkable optical properties and ultra-high Q-factors (up to $10^{11}$), especially in fluoride crystals [3]. Si bulk WGM resonators [4] have been recently shown with $Q=10^9$. WGM microresonators have been exploited for both fundamental studies and practical applications [5] because of their unique properties of long cavity lifetime and small mode volumes. The strongly enhanced light-matter interaction allows implementing nonlinear and quantum sources [6] and demonstrating key functionalities like frequency combs [7] and high-purity radiofrequency signal generation [8], optomechanical oscillations [9], stimulated inelastic scattering [10], laser stabilization [11], and biosensing [12]. The flexibility of the techniques involved in their fabrication, allows choosing within a wide range of materials, making WGM resonators a unique platform in terms of performance and versatility.

A critical point across all the aforementioned applications, is the implementation of an efficient, controllable and robust coupling of the light to the resonator modes [13]. The best approach so far relies on phase-matched evanescent field coupling and requires some overlap of the evanescent field of the WGM with the evanescent field of the coupler. Most state-of-the-art demonstrations rely on inherently fragile thin fiber tapers, with few-microns diameters, which are ideal only for lab demonstrations. More robust approaches are based on angle cleaved fibers [14] and integrated waveguides [15]. Prism-coupling [16] is a particularly versatile technique but relies on free space optics and requires the beam to spatially match the resonator mode, which can be a critical and limiting factor in several circumstances.

There are only a couple of examples in the literature where silicon photonic waveguides are coupled to high-index bulk resonators: chalcogenide spheres [17] and lithium niobate disks [18]. However, efficient and robust coupling to bulk resonators made with low refractive index materials, e.g. silica and alkaline earth fluorides, remains an open challenge due to the huge phase mismatch between high effective index silicon waveguide and low effective index resonators. Coupling of silica microtoroid and silicon waveguide has been reported based on a suspended silicon photonic crystal membrane [19]. Yet, this strategy exhibits a compromised mechanical stability (suspended membrane) and a narrow bandwidth limited to a small wavelength range close to the photonic crystal band edge. Low refractive index ultra-high Q WGM resonators made from alkaline earth fluorides were only recently coupled to a low refractive index suspended silica waveguide [20]. Recently, coupling of on-chip silica microtoroid and integrated silicon nitride waveguides has been demonstrated [21]. However, this solution requires complex fabrication process and yields coupling only to one specific type of microresonator while still requiring coupling between silicon nitride and Si waveguides to exploit the advantages of Si photonics.

None of the above coupling techniques allows efficient coupling to different kinds of resonators, limiting their capability to exploit the wide range of materials and optical properties provided by bulk WGM resonators. Here, we propose and demonstrate a new type of universal coupling approach able to interconnect integrated silicon waveguides and a wide range of bulk WGM resonators with largely disparate sizes (300 μm – 3.6 mm diameter) and refractive indices (1.42-2.21). The proposed approach utilizes the unique flexibility of subwavelength grating (SWG) waveguides to shape the optical field distribution and wavevector of guided modes. We implement a tapered geometry, presented in Fig. 1(a), where the SWG geometry is continuously varied to allow efficient coupling to different kinds of resonators. These results open a new path for the development of universal integrated coupler able to combine silicon photonics with almost any bulk resonator.

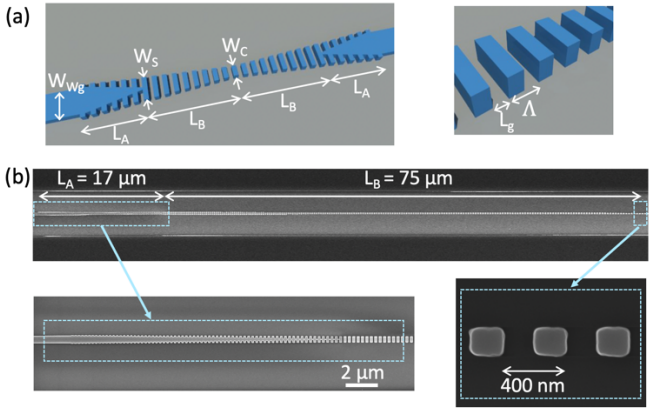

**Fig. 1**. (a) Schematic diagram of proposed SWG metamaterial tapered waveguide designed for coupling light to bulk WGM resonators. (b) Scanning electron microscope image of SWG metamaterial coupler.

SWG waveguides [XX(*to be completed*)] provide precise lithography control of the optical field distribution non feasible with strip and rib geometries, with lower propagation loss and wider bandwidth than photonic crystals. SWGs have allowed the demonstration of plethora of photonic devices with unprecedented performance, including fiber-chip couplers [22, 23], biosensors [24, 25], wavelength filters [26], waveguide crossings [27] and metamaterial lenses [28]. However, they have not been exploited yet as WGM microresonators couplers.

Figure 1(a) shows the schematic view of the proposed SWG coupler. Silicon thickness is t = 220 nm. At both ends, the coupler is connected to conventional strip waveguides ($W_{wg}$ = 500 nm) by adiabatic transitions of length $L_A$ = 17 μm, similar to those presented in [29]. The SWG taper has a length of $L_B$, with a linear variation of the period ($\Lambda$) and gap length ($L_g$). At the beginning of the taper, the SWG waveguide has a width of $W_S$ = 400 nm, a period of $\Lambda$ = 270 nm and a gap length of $L_g$ = 100 nm. At the end of the taper the SWG waveguide has a period of $\Lambda$ = 400 nm and a gap length of $L_g$ = 200 nm. The waveguide width at the end of the taper, $W_C$, is varied between 200 nm and 450 nm to implement different effective index profiles. By increasing the gap length ($L_g$) and narrowing the waveguide width ($W_C$), we expand the optical mode and reduce its effective index. The geometry of the coupler has been optimized to achieve adiabatic mode transformation with low insertion loss and negligible excitation of higher order modes for transverse-electric (TE) polarized mode near 1550 nm wavelength.

We have fabricated 24 different SWG couplers. The final taper width, $W_C$, has been varied between 200 nm and 450 nm with a step of 50 nm. For each value of $W_C$, four different taper lengths, of $L_B$ = 75 μm, 150 μm, 300 μm and 450 μm, have been implemented. The devices are fabricated by electron-beam lithography and reactive ion etching using a 220-nm-thick single crystal silicon layer of a silicon-on-insulator wafer (with a 3-μm-thick bottom silica layer). The samples are then spin coated with a PMMA layer (upper cladding) of 300 nm thickness with a refractive index of 1.5. Figure 1(b) shows scanning electron microscope (SEM) images of the coupler with $W_C$ = 200 nm and $L_B$ = 75 μm.

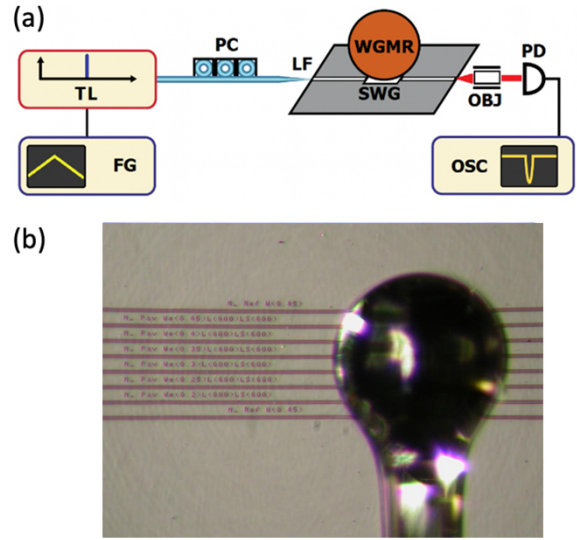

**Fig. 2**. (a) Simplified schematic of the setup. Function generator (FG), Tunable Laser (TL), Polarization Controller (PC), Lensed Fiber (LF), Sub-Wavelength Grating waveguide (SWG), 10X Objective (OBJ), Photodetector (PD), Oscilloscope (OSC). (b) Optical microscope image of a WGM microsphere resonator placed above the silicon chip containing a set SWG waveguides. The resonator is centered above and in the middle of the SWG tapered waveguide coupler.

In order to test the novel SWG couplers, different WGM resonators made from $CaF_2$ ($n_{CaF2}$=1.42), $MgF_2$ ($n_{MgF2}$=1.37), silica ($n_{SiO2}$=1.44), and $LiNbO_3$ ($n_{LiNbO3}$=2.21) were fabricated. We have employed arc-discharge melting technique for manufacturing silica microspheres (of diameter between 20-400 μm) [30, 31]. Polishing technique with diamond suspensions has been used for $CaF_2$, $MgF_2$, silica and $LiNbO_3$ disks (of diameter between 2 and 5 mm and thickness of 0.5 mm) [15]. A sketch of the experimental setup is shown in Fig. 2(a). A continuous-wave tunable external-cavity laser operating around 1550nm with a linewidth of 300 kHz is modulated, applying a linear ramp through a function generator, and coupled to the silicon waveguide in the chip via a lensed single mode fiber with 2 μm Gaussian beam waist. The output is collected through a 10× objective and a photodiode and analyzed using an oscilloscope. The overall transmission is ~1%, mainly due to rather high fiber-waveguide coupling losses combined with adapter and tapered SWG losses. Figure 2(b) shows the relative position between a WGM microsphere resonator and a set SWG waveguides.

We start with the study of the silica microsphere. As shown in the transmission spectrum of Fig. 3, silica microspheres can be critically coupled to tapered SWGs with $W_C$ = 400 nm. The measured coupling efficiency is CE ~ 99%, and the maximum quality factor is Q ~ 2 ×$10^7$. Decreasing the width $W_C$, results in a reduction of the resonance contrast, down to 40% for $W_C$ = 200 nm. No coupling difference has been observed changing the SWG coupler lengths ($L_B$ = 300, 150 and 75 μm). The coupling performance has been evaluated when the microsphere is in contact with the chip surface and when there is small separation. In contact, the Q-factor decreases to ~5 x $10^6$, with no variation of coupling efficiency.

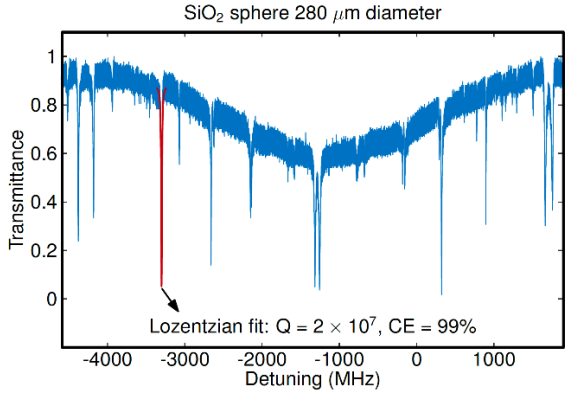

**Fig. 3.** Transmission spectrum of a SWG waveguide coupled to a silica microsphere of 280 μm in diameter. Maximum measured coupling efficiency (CE, defined as the resonance depth or contrast) is 99%. The blue line is the zero-transmission baseline obtained minimizing the transmission by changing the polarization state of the light injected in the silicon waveguide. The red line represents the Lorentzian fit of a resonance dip.

Then, we study the coupling with disks made of different materials ($LiNbO_3$, $SiO_2$, $CaF_2$, and $MgF_2$) with millimeter-scale radius. Z-cut lithium niobate disk, with high-refractive index of $n_{LiNbO3}$=2.21, is critically coupled to SWG taper with $W_C$ = 450 nm, reaching a Q-factor of ~ 8 ×$10^6$, as shown in Fig. 4(a). No differences in coupling efficiency have been observed when changing the SWG coupler length $L_C$ between 450 μm and 75 μm. For tapered SWG with $W_C$ = 400 and 350 nm the coupling efficiency is reduced to CE = 40%-50%. For $W_C$ = 300, 250 and 200 nm, CE drops below 5%. CE has been evaluated in no-contact configuration since, in-contact, the transmission is affected by the disk presence, completely extracting light from SWGs.

Silica disk, with lower refractive index of $n_{SiO2}$=1.44, is critically coupled to tapered SWGs with central width of $W_C$ = 350, 300, 250 and 200 nm. As shown in Fig. 4(b), the measured Q-factor reaches up to Q ~ 1.7 x $10^7$. No critical coupling has been obtained for $W_C$ = 400 nm (up to 60% CE). No differences have been observed changing the SWG coupler length (i.e. for $L_C$ = 450, 300, 150 and 75 μm). Compared with those of $SiO_2$ microspheres, these results indicate that larger resonators are better coupled to a wider range of SWG geometries.

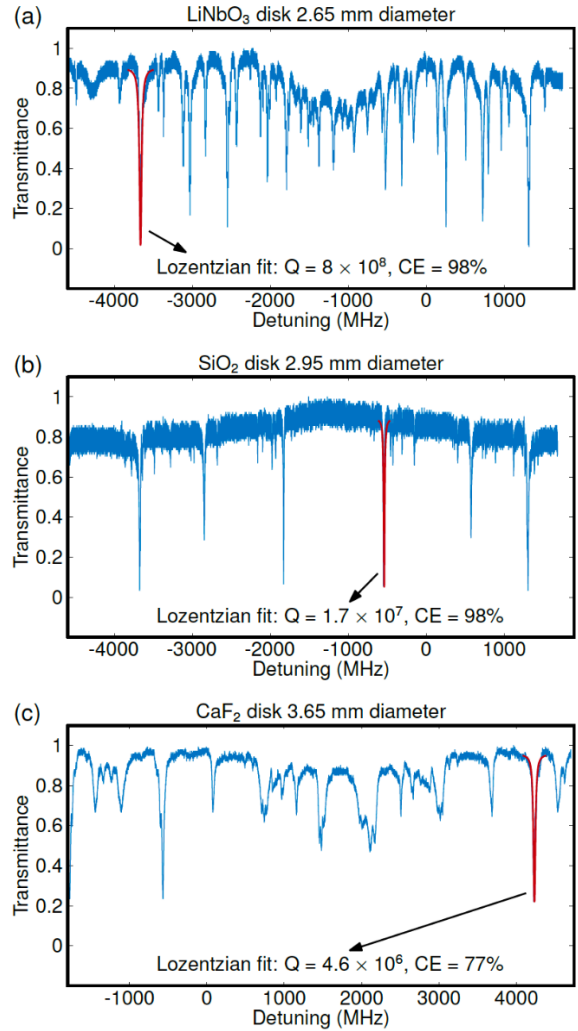

**Fig. 4.** Transmission spectrum of a SWG waveguide coupled to: (a) $LiNbO_3$ disk of 2.65 mm in diameter. Maximum measured coupling efficiency is 98%. (b) a silica disk of 2.95 mm in diameter. Maximum measured coupling efficiency is 98%. (c) a $CaF_2$ disk of 3.65 mm in diameter. Maximum measured coupling efficiency is 77%.

Calcium fluoride disk, with lower refractive index of $n_{CaF2}$=1.42, yields high coupling efficiency CE = 77 %, with a Q-factor of ~ 4.7 x $10^6$, for $W_C$ = 250 nm and $L_B$ = 150 μm. However, critical coupling is not achieved with any of the SWG couplers. Magnesium fluoride disk, with the lowest refractive index considered here ($n_{CaF2}$=1.42) yields a comparatively poor coupling efficiency, CE<5%, achieved only for $W_C$ = 250 nm and 200 nm and taper length of 75 μm. Q factors do not exceed ~$10^5$. These results may suggest that coupling to resonators with very low refractive index would require narrower waveguide width or shorter taper lengths.

Figure 5 summarizes the best results showing optimized SWG width maximizing the coupling efficiency for bulk WGM resonators we investigated.

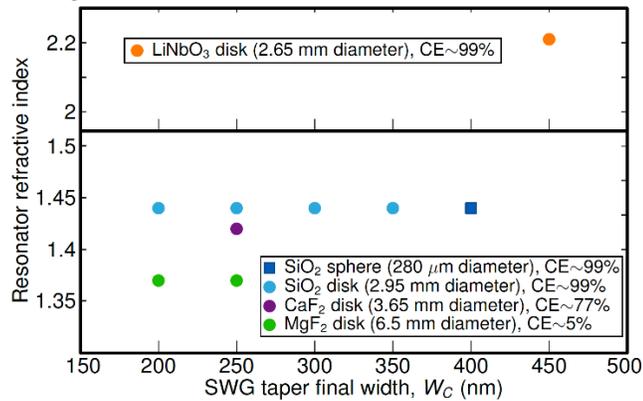

**Fig. 5.** Optimum SWG taper width, $W_C$, maximizing the measured coupling efficiency for the bulk WGM resonators studied here.

In conclusion, we have shown that SWG waveguide-based couplers provide unique flexibility to combine a wide range of bulk WGM microresonators with nanometric-thick Si waveguides, achieving high coupling efficiency. The proposed structure represents a proof-of-concept demonstration of a universal integrated photonic coupling approach, that can be seamlessly fabricated with standard Si technology, opening a new route to exploit a wide range of optical properties exalted by ultra-high Q resonances. The coupling mechanism presented here cannot be fully described by conventional effective index matching theory. The effective index in the SWG couplers vary between 2.8 and 1.6. Hence, only coupling to $LiNbO_2$ ($n_{LiNbO3}$=2.21) could be described by index matching theory. Fully describing the coupling to low index bulk resonators made of silica and calcium fluoride may require the development of a new theoretical framework where the phase-gradient effect [32] could play a key role. Still, further investigations are required. Nevertheless, these unprecedented results open a new path with great potential for applications requiring high resolution spectral filtering or light recirculation, for instance in in sensing, microwave-photonics, and quantum photonics, to name a few.

**Funding.** National Science Foundation (NSF) (1263236, 0968895, 1102301); The 863 Program (2013AA014402). Agence Nationale de la Recherche, BRIGHT ANR-18-CE24-0023-01.

**Acknowledgment**. We thank Prof. F. Cosi for his help with the experiment.